\documentclass[twocolumn,showpacs]{revtex4}

\usepackage{graphicx}
\usepackage{dcolumn}
\usepackage{amsmath}
\usepackage{epsfig}
\usepackage{amssymb}

\newtheorem{theorem}{Theorem}

\newcommand{\ket}[1]{\left\vert#1\right\rangle}
\newcommand{\bra}[1]{\left\langle#1\right\vert}

\newcommand{\dmat}[2]{\ket{#1}\!\!\bra{#2}}

\newcommand{\beq}{\begin{equation}}
\newcommand{\eeq}{\end{equation}}
\newcommand{\bea}{\begin{eqnarray}}
\newcommand{\eea}{\end{eqnarray}}

\makeatletter
\def\btt#1{\texttt{\@backslashchar#1}}
\DeclareRobustCommand\bblash{\btt{\@backslashchar}}
\makeatother
\def\Bid{{\mathchoice {\rm {1\mskip-4.5mu l}} {\rm
{1\mskip-4.5mu l}} {\rm {1\mskip-3.8mu l}} {\rm {1\mskip-4.3mu l}}}}

\begin{document}

\title{All maps equivalent to a given map, completely positive or not}
\author{Yong-Cheng Ou$^{1}$} 
\author{Mark S. Byrd$^{1,2}$} 
\affiliation{$^1$Physics Department,
Southern Illinois University,
Carbondale, Illinois 62901}
\affiliation{$^2$Computer Science Department,
Southern Illinois University,
Carbondale, Illinois 62901}

\date{\today}

\begin{abstract}
A dynamical map is a map which takes one density operator to another.
Such a map can be written in an operator-sum
representation (OSR) using a spectral decompositon.  The method of the
construction applies to more general maps which need not be completely
positive.  The OSR not unique; there is a freedom 
to choose the set of operators in the OSR differently, yet still
obtain the same map.  Here we identify all maps which are 
equivalent to a given map.  Whereas the freedom for completely
positive maps is unitary, the freedom for maps which are not
necessarily completely positive is pseudo-unitary.  
\end{abstract}


\maketitle


\section{Introduction}

A quantum system may undergo unitary evolution in the simple case
that the system is closed.  However, many practical
experiments involve systems for which this is not a good
approximation.  In these cases, the evolution is an open-system
evolution where the external influences can be very important.  When
these systems are being considered for quantum information processing
and/or transfer, understanding the external influence and modeling it 
is often a necessary part of the process of evaluating experiments.  

The need for describing such evolution of a particular system using 
dynamical maps was identified by Sudarshan, Mathews, and Rau 
\cite{Sudarshan:61}.  Their work provides a mapping from one density  
operator to another which can be used to describe 
a quite general open-system evolution.  
Some time later Kraus \cite{Kraus:83} restricted
consideration to maps which are completely positive, a useful but not
necessary assumption for open system evolution. The importance of the  
complete positivity assumption has recently been discussed in the 
literature
\cite{Pechukas:94,Pechukas+Alicki:95,Jordan:04,Shaji/Sudarshan:05,Shabani/Lidar:09a}.
In particular, it has recently been shown that vanishing quantum
discord \cite{Oll/Zurek:01} is sufficient \cite{Cesar/etal:08} (see
also \cite{Pechukas+Alicki:95}) and
necessary \cite{Shabani/Lidar:09a} for completely positive maps.
Furthermore, the conditions an initial state must satisfy in order for
a map to be positive were also found in Rev.~\cite{Shabani/Lidar:09b}.

In the case of a completely positive map, it is well-known that 
a unitary degree of freedom exists in the OSR.  This freedom for
completely positive maps is noted and 
used by Nielsen and Chuang 
\cite{Nielsen/Chuang:book} for error correction, modeling, and other
applications to quantum information processing.  Most
notably perhaps is the application to quantum error correcting codes.  
In that case it was shown that the freedom can be used to model errors
in a very useful way by choosing a complete basis for errors which
acts as a basis for the operators in the operator-sum representation
(OSR).  Recently it was shown that 
complete positivity is not a necessary assumption for quantum error
correcting codes \cite{Shabani/Lidar:09b}.

In this article, we identify the pseudo-unitary degree of freedom in
the OSR for maps that are not necessarily completely positive.  We
first provide some background and the origin of the OSR.  We then
review the unitary degree of freedom for completely positive maps.
Finally, we show that the freedom for the more general case of maps
which are not necessarily completely positive, before discussing the
implications of this freedom.


\section{Background}

In 1961, Sudarshan, Mathews, and Rau (SMR) described what they called
``dynamical maps'' \cite{Sudarshan:61}.  These are maps from one
density matrix (or density operator) 
to another with no other restrictions.  The authors
were able to arrive at a form for the map which is quite
general and provides conditions for the map to be positive.  We
provide their argument here as a basis for what follows. 


\subsection{Dynamical Maps and the SMR Decomposition}
\label{sec:osr}

Following the treatment of Sudarshan, Mathews, and Rau, let us
consider a quite general mapping from one Hermitian matrix to another 
of the form
\begin{equation}
\label{eq:Asnosub}
\rho^\prime = A \rho,
\end{equation}
or more explicitly
\begin{equation}
\label{eq:As1}
\rho^\prime_{r^\prime s^\prime} = A_{r^\prime s^\prime,rs} \rho_{rs}.
\end{equation}
It is apparent that this is a very general linear map $A$ 
maps elements of $\rho$ to elements of another operator 
$\rho^\prime$.  For this reason it is sometimes referred to as a
superoperator.  Now, we recall that the density matrix is
required to be Hermitian, positive semi-definite, and have
trace one.  Respectively, we write
\begin{equation}
\label{eq:hermiticitypositivitynormalization}
\rho = \rho^\dagger, \;\;\; \rho \geq 0,\;\;\; \mbox{Tr}\rho = 1.
\end{equation}
These ensure a valid probability interpretation of the density matrix.
One can show that, given the properties specified in 
Eq.~(\ref{eq:hermiticitypositivitynormalization}), the mapping
$A$ must have the following properties:
\begin{equation}
A_{sr,s^\prime r^\prime} = (A_{rs,r^\prime s^\prime})^*,
\end{equation}
\begin{equation}
x_r^*x_sA_{sr,s^\prime r^\prime}y_{s^\prime}y_{r^\prime} \geq 0,
\end{equation}
and
\begin{equation}
A_{rr,s^\prime r^\prime} = \delta_{s^\prime r^\prime}.
\end{equation}
As SMR, we introduce a new matrix $B$ which is related to
$A$ by relabeling,
\begin{equation}
B_{rr^\prime,s s^\prime}\equiv A_{sr,s^\prime r^\prime},
\end{equation}
with the following properties:
\begin{equation}
B_{rr^\prime,s s^\prime}=(B_{ss^\prime,r r^\prime})^*,
\end{equation}
\begin{equation}
z^*_{rr^\prime}B_{rr^\prime,s s^\prime}z_{ss^\prime} \geq 0,
\end{equation}
and
\begin{equation}
B_{rr^\prime,r s^\prime} = \delta_{r^\prime s^\prime}.
\end{equation}
Note that $B$ can be considered
a Hermitian matrix, and as such, it is diagonalizable, i.e. it has a
spectral decomposition, or eigenvector decomposition.  Now, noting
(\ref{eq:Asnosub}),  can be written $\rho^\prime = B\rho$ and letting
$\eta_k^\prime$ be the eigenvalues of $B$ the spectral decomposition
is 
\begin{equation}
\label{eq:OSdecompcomponents}
B_{rr^\prime,s s^\prime} = \sum_k
               \eta_k^\prime
               \xi^{(k)}_{rr^\prime}\xi^{(k)\dagger}_{s^\prime s}.
\end{equation}
This is also sometimes written with the component indices supressed
as follows:
\begin{equation}
\label{eq:OSdecomp}
B = \sum_k \eta_k^\prime C_k^{\prime{\phantom{\dagger}}} C_k^{\prime\dagger},
\end{equation}
where $(C_k)_{rr^\prime}=\xi^{(k)}_{rr^\prime}$, or more succinctly as 
\begin{equation}
B = \sum_k \eta_k^\prime \dmat{k^\prime}{k^\prime},
\end{equation}
where $C_k^\prime = \ket{k^\prime}$ is an eigenvector of $B$.  
(For more detail see for example \cite{Jordan:04}.)   The $C_k$, as
the density operator itself, may be written as a matrix or a vector.
Then Eq.~(\ref{eq:Asnosub}) may be considered a linear map of a vector
to another vector via matrix multiplication.  Clearly both pictures
are consistent with Eq.~(\ref{eq:As1}).

If we assume that the $\eta^\prime_k$ are all positive, we
may absorb them into the $C$'s to arrive at the familiar form of the
operator-sum decomposition:
\begin{equation}
\label{eq:dmatstsmr}
\rho^\prime = \sum_k
               A_{k}^{\phantom{\dagger}}\rho A_{k}^\dagger,
\end{equation}
where $A_k = \sqrt{\eta_k^\prime} C_k^\prime$ \cite{Krausnote}.

It is interesting to note that this ``spectral decomposition'' or 
``eigenvector decomposition'' of 
the map gives a minimal decomposition of the map as defined and used in
Refs.~\cite{Timoney:00} and \cite{Jam:04} 
since the eigenvectors are linearly independent.  It follows that
the operators are linearly independent.  


\subsection{Hermitian Preserving Maps}

Here we consider the case where the map does not necessarily
correspond to a trace-perserving, completely positive map.  We will 
consider maps which are hermiticity-preserving, i.e. they take
hermitian matrices to hermitian matrices.  (They are also sometimes
called Hermitian-preserving.)  Thus a general map $\Phi$ can be
expressed as
\beq
\label{eq:OSmapdecomp}
\Phi(\rho) = \sum_k
               \eta_k^\prime C_k^{\prime{\phantom{\dagger}}}\rho C_k^{\prime\dagger},
\eeq
and again, when the $\eta_k^\prime$ are all positive, 
$\Phi$ can be expressed as
\beq
\label{eq:stsmr}
\Phi(\rho) = \sum_k
               A_{k}^{\phantom{\dagger}}\rho A_{k}^\dagger.
\eeq
Note that if the $\eta_k^\prime $ are not all positve,
then we may take $\eta_k = (\pm 1)$ and the square-roots of the
postive magnitudes, $\sqrt{|\eta_k '|}$ may be
absorbed into the $C_k^\prime$.  (The $\eta_k^\prime$ are real since $B$ is
Hermitian.)  Therefore, we let $C_k = \sqrt{|\eta_k '|}C_k^\prime$ and
express Eq.~(\ref{eq:OSmapdecomp}) as
\begin{eqnarray}
\label{eq:2posmaps}
\Phi(\rho) &=& \sum_{k=1}^{p+q} \eta_k C_k^{\phantom{\dagger}}\rho C_k^\dagger,
                \nonumber \\
           &=& \sum_{k=0}^p C_k^{\phantom{\dagger}}\rho
              C_k^\dagger -  \sum_{k=p+1}^{p+q} C_k^{\phantom{\dagger}}\rho C_k^\dagger,
\end{eqnarray}
assuming there are a total of $p+q$ terms in the sum with $\eta_k
=+1,$ for $k=1,..., p$ and $\eta_k=-1,$ for $k=p+1,...,p+q$.  
This is an expression of the fact that any
Hermitcity-reserving map can be written as the difference between two
completely postive maps \cite{Sudarshan/Shaji:03}.


\section{Unitary Freedom in the OSR}

The description of the dynamical map is not unique.  It can be
represented by the set of $C_k$ corresponding to the
eigendecomposition of the map $B$, but there are many other
representations.  In this section we find
this freedom after reviewing the case for completely positive maps. 
For completely positive maps, we reiterate that a theorem describing
the freedom, examples, and uses can be found in 
Ref.~\cite{Nielsen/Chuang:book} although our presentation
differs somewhat from theirs.


\subsection{Unitary Freedom for Completely Positive Maps}

A completely positive map can always be written in the form given in
Eq.~(\ref{eq:stsmr}).  If we then let a new set be given by
$A^\prime_j = \sum_i u_{ji}A_i$, then
\begin{equation}
\label{eq:cpequiv}
\Phi^\prime(\rho) = \sum_k A_k^\prime \rho A^{\prime\dagger}_k
                  =\sum_{ijk} u_{jk}A_{k}^{\phantom{\dagger}}\rho u_{ji}^*A_{i}^\dagger.
\end{equation}
which is the same as $\Phi$ if and only if
\beq
\label{eq:unitarycond}
\sum_j u_{jk}u_{ji}^* = \delta_{ik}.
\eeq  
Eq.~(\ref{eq:unitarycond}) is the 
condition for the set of numbers $u_{ij}$ to form a unitary matrix,
$U^\dagger U = \Bid$.  
However, there is no restriction on the number of elements in the set
$\{A_i\}$ compared to the number of elements in the set
$\{A_k^\prime\}$ except that the matrix
between them satisfy the condition Eq.~(\ref{eq:unitarycond}).  Due to
this, the matrix is sometimes referred to as left-unitary 
which is an $n\times m$ matrix $T$ such that
$T^\dagger T =\Bid_m$.  Otherwise one may append zero operators to the
shorter smaller set and make the matrix $U$ a 
square unitary matrix \cite{Nielsen/Chuang:book}.


\subsection{Pseudo-unitary freedom for Hermiticity-preserving maps}

Now let us consider the map $\Phi(\rho) = \sum_j \eta_j  C_j
\rho C_j^\dagger$ and introduce a set of operators $D_j$ corresponding
to the map $\Phi '(\rho) = \sum_j \eta_j D_j \rho D_j^\dagger$.  As
stated above, we may take $\eta_j = \pm 1$.  We can choose 
the number of operators to be the same by appending
zero operators to the shorter list.  
This enables the number of $-1$ and $+1$ to be chosen to be the same.  
Furthermore, we will order the
set of $\eta_j$ such that the first $p$ are $+1$ and the next $q$ are
$-1$.  

As stated in the title of this section, the freedom in the
operator-sum representation for maps which are not necessarily
completely positive is a pseudo-unitary degree of freedom.  By this we
mean the freedom is described by the group $U(p,q)$.  This
group is often called a pseudo-unitary group due to its relation to the
unitary group and it is a metric-preserving group with the signature
of the metric determined by the integers $p,q$.  See
for example (\cite{Gilmore:book}, pages 45, 197),
(\cite{Cornwell:84II}, page 392), (\cite{Wybourne}, page 12), or 
(\cite{Helgason:DS}, page 444). 

Let $\eta$ be an $N\times N$ diagonal matrix with
the first $p$ entries $+1$, the next $q$ entries $-1$, and $N=p+q$.  Then
for all $U\in U(p,q)$, 
\begin{equation}
\label{eq:upqcond}
U^\dagger \eta = \eta U^{-1}.
\end{equation}
We may express the matrix $\eta$ as a diagonal matrix with the matrix
elements being $\eta_k$, $\eta_k = +1$, for $k = 1, ..., p$ and
$\eta_k = -1$, for $k=p+1, ..., p+q=N$.  Alternatively, we may express
the matrix $\eta$ using elements $(\eta)_{kl} = \eta_k \delta_{kl}$.
This is clearly a diagonal matrix since the elements are zero if
$k\neq l$.  Furthermore, the first $p$
entries along the diagonal are $+1$ and the next $q$ are $-1$.  Let the
elements of the matrix $U$ be given by $u_{ij}$ and those of 
$U^\dagger$ be $u_{ji}^*$.  Then the Eq.~(\ref{eq:upqcond}) can be
written as $U^\dagger \eta U = \eta$, or since $\eta^2 = \Bid$, 
$U \eta U^\dagger = \eta$.  In components, this can be written as 
\begin{equation}
\sum_{jk} u_{ij} \eta_j \delta_{jk} u_{lk}^* = \eta_i\delta_{il}. 
\end{equation}
Having establihsed this property for elements of the group $U(p,q)$,
the following theorem may now be stated.

\begin{theorem}{Pseudo-unitary freedom:}
Suppose $\{C_1,C_2, ..., C_n\}$ and  $\{D_1,D_2, ..., D_m\}$, 
are operation elements giving rise to quantum operations (maps) 
$\Phi$ and $\Phi '$ respectively.  Explicitly, 
\begin{equation}
\Phi = \sum_i\gamma_iC_iC_i^\dagger
\end{equation}
and 
\begin{equation}
\Phi^\prime = \sum_j \mu_j D_jD_j^\dagger,
\end{equation}
where each $\gamma_i$ and each $\mu_j$ is $\pm 1$ and ordered as
above, with all $+1$ eigenvalues first.  Furthermore, we can always
take $\gamma_i = \mu_i$ with zero-valued $C_i$ or zero-valued $D_j$
appended to the shorter list for the $+1$ $(-1)$ eigenvalue.  
Then $\Phi=\Phi^\prime$ if and only if 
\begin{equation} \label{e123}
D_j =\sum_{i} u_{ji}C_{i},
\end{equation}
where the numbers $u_{ij}$ form a $p+q$ by $p+q$ 
matrix in $U(p,q)$.  
\end{theorem}

{\it Proof:} We first consider whether the condition is necessary and
use the notation $C_i=\ket{i}$, $D_i=\ket{j}$.  Suppose that 
\begin{equation}
\Phi =\Phi^\prime.
\end{equation}
(Or, if one would like to display the argument explicitly, 
$\Phi(\rho) =\Phi^\prime(\rho)$.)  For a general map $\Phi$, there
exists a corresponding $B$ matrix (see Sec.~\ref{sec:osr}) such
that $\Phi=B$ (i.e. $\Phi (\rho) = B \rho$).  $B$ has an eigenvector
decomposition $B=\sum_{k^\prime} \lambda_k^\prime
\dmat{k^\prime}{k^\prime}$ where the set of
$\ket{k^\prime}$ are linearly independent since they are orthogonal. 
This follows
from the fact that the eigenvectors can be chosen orthogonal. Now 
$|k\rangle=\sqrt{|\lambda_k^\prime|}\;
|k^\prime\rangle$.  These vectors are clearly also orthogonal and thus
linearly independent if the $\ket{k^\prime}$ are.  Then $B$ can be
re-expressed as 
$B=\sum_k \eta_k \dmat{k}{k}$ with the first $p$ eigenvalues 
$\eta_k=+1, \; k= 1,...,p$ and the next 
$q$ eigenvalues $\eta_k=-1\;, k=p+1,...,p+q$.  This gives 
\begin {equation}
B =\sum_k \eta_k \dmat{k}{k} 
  = \sum_{k=1}^{p} \dmat{k}{k}-\sum_{k=p+1}^{p+q} \dmat{k}{k},
\end{equation}
which is an eigenvector decomposition of the map $\Phi$,
Eq.~(\ref{eq:OSdecomp}).  
Now, let us consider another decomposition of $B$ corresponding to the
set of $C_i$, $B=\sum_{i} \gamma_i \dmat{i}{i}$.  Each $|i\rangle$
can be written as a linear combination of the $|k\rangle$,
$|{i}\rangle =\sum_{k} w_{ik}|{k}\rangle$. (See for example 
Ref.~\cite{Nielsen/Chuang:book}, page 104.) 
Given $\Phi=B$ 
\begin{equation}
  \sum_k \eta_k  \dmat{k}{k} = \sum_{kl}\left(\sum_i 
                \gamma_i w_{ik}w^*_{il}\right)\dmat{k}{l}.  
\end{equation}
Since the $\ket{k}$ are linearly independent, it is clear that this
can only happen if 
\begin{equation} \label {e1}
\sum_i\gamma_i w_{ik}w^*_{il} = \delta_{kl}\eta_k.
\end{equation}
We may always take $\eta_i = \gamma_i$ by appending the shorter list
of vectors ($\{\ket{i}\}$ or $\{\ket{k}\}$) with zero vectors.  
This will ensure the matrices $\gamma$ with elements
$\delta_{ij}\gamma_i$ and $\eta$ with elements $\delta_{kj}\eta_k$ are
equal.  Furthermore,  $w$ can then be taken to be square with $\ket{i}
= \sum_k w_{ik}\ket{k}$.  The condition, Eq.~(\ref{e1}), can then be
written as 
\begin{equation}
w^\dagger \eta w = \eta,
\end{equation}
which is the condition for the matrix $w$ to be in $U(p,q)$.  
Now, we can use the same argument, with $B=\Phi^\prime$ and 
$v_{jk}$ such that  $\ket{j} = \sum_k v_{jk}\ket{k}$, to show
\begin{equation} \label{a7}
v^\dagger \eta v =\eta. 
\end{equation}
Since each of these two are related to the same expression for $B$
using elements of $U(p,q)$ which is a group, then the linear
transformation which takes the $C_i$ to the $D_j$, 
$u=vw^{-1}$ is in $U(p,q)$.  

Next, we consider whether $u\in U(p,q)$ will imply that
$\Phi=\Phi^\prime$, i.e., if the condition is sufficient.  
This is straight-foward algebra. 
Given Eq.(\ref{e123}) 
\begin{eqnarray} \nonumber
  \Phi^\prime(\rho) &=& \sum_j \mu_j D_j \rho D_j^{\dag}\\ \nonumber
  &=& \sum_{lkj} \mu_j u_{jl}u^*_{jk}C_l \rho C_k^{\dag}\\ \nonumber
   &=& \sum_{lk}\left(\sum_j \mu_ju_{jl}u^*_{jk}\right)C_l \rho
   C_{k}^{\dag} \\ \nonumber
     &=& \sum_{l}\gamma_l \delta_{lk} C_l \rho C_{k}^{\dag} \\ \nonumber
            &=& \Phi (\rho),
\end{eqnarray}
which shows that the two sets of operators $C_j$ and $D_j$ related 
by a pseudo-unitary matrix $u$ will yield the same map. $\square$


\section{Conclusions}

The unitary degree of freedom in the operator-sum representation of
quantum maps has multiple uses including applications in quantum error
prevention schemes since it essentially provides the set of physical
operators producing a given map.  
With the recent extensive discussions in the
literature concerning maps which are not completely positive, and the
extension of quantum error correction to maps which are not completely
positive, we believe it is important to have an 
extension of the unitary degree of freedom for completely
positive maps to the cases when the map is not necessarily completely
positive.  We have done that here by providing the pseudo-unitary freedom
for any Hermitian-preserving map.  

It is natural to ask which maps are genuinely different in the
sense that they are not equivalent.  In other words, if we consider
the unitary degree of freedom to be a symmetry of the system, what is
unique to two different maps?  This can, in 
principle, be determined from the work here and 
Ref.~\cite{Sudarshan/Shaji:03} where the authors parameterized the
space of positive maps.  Also we see that the minimal decomposition
provided by the spectral decomposition is very important.  The map $B$
is unique  \cite{Sudarshan/Shaji:03} and provides an almost canonical
form \cite{cannote} which has a set number of positive and negative 
operators when written as the sum of two completely 
positive maps.  These important issues should be addressed in future work.


\section*{Acknowledgments}

This material is based upon work supported by NSF-Grant No. 0545798 to MSB.
We acknowledge C. Allen Bishop for helpful discussions.




\begin{thebibliography}{20}
\bibitem[{can()}]{cannote}
\bibinfo{note}{{The eigenvalue decomposition is unique if there are no
  degeneracies and we choose an orthonormal basis.}}
\end{thebibliography}
\end{document}